# A Twist on Broken U(3)xU(3) Supersymmetry

Scott Chapman
schapman@chapman.edu
*Chapman University, One University Drive, Orange, CA 92866*
(Dated: May 1, 2020)

**Abstract**
What symmetry breaking would be required for gauginos from a supersymmetric theory to behave like left-handed quarks of the Standard Model?  Starting with a supersymmetric SU(3)xSU(3)xU(1)xU(1) gauge theory, the 18 adjoint-representation gauginos are replaced with 2 families of 9 gauginos in the (3,3*) representation of the group.  After this explicit breaking of supersymmetry, two-loop quadratic divergences still cancel at a unification scale.  Coupling constant unification is supported by deriving the theory from an SU(3)xSU(3)xSU(3)xSU(3) Grand Unified Theory (GUT).  Sin^2 of the Weinberg angle for the GUT is 1/4 rather than 3/8, leading to a lower unification scale than usually contemplated, ~10^9 GeV.  After spontaneous gauge symmetry breaking to SU(3)xSU(2)xU(1), the theory reproduces the main features of the Standard Model for two families of quarks and leptons, with gauginos playing the role of left-handed quarks and sleptons playing the role of the Higgs boson.  An extension to the theory is sketched that incorporates the third family of quarks and leptons.

**Introduction:**

Non-supersymmetric theories typically have quadratic divergences that lead to the Hierarchy (or fine tuning) problem [1], which could be simply stated as follows:  What causes an infinite string of terms of order $10^{38}$ GeV^2 to sum up to a Higgs boson mass squared of $\sim 10^4$ GeV^2?  In supersymmetric theories, quadratic divergences are completely (or mostly) cancelled, thereby solving (or greatly lessening) the Hierarchy problem.  However, most supersymmetric theories that describe the observed particle spectrum have a basic assumption in common:  For each particle that has been observed experimentally (quark, lepton, gauge boson, Higgs boson), there is a supersymmetric partner particle that has not yet been observed (squark, slepton, gaugino, Higgsino).  As experiments get to higher and higher energies without detecting any of these partners, some have begun to doubt whether supersymmetry (SUSY) exists in Nature.

Others have taken a new look at an old question:  What if some of the particles that we currently observe are supersymmetric partners with each other?  For example, what if the Higgs boson is the "slepton" partner of the electron-neutrino doublet [2-5]? Or what if certain components of the Higgs boson are in the same SUSY representation as the W and Z bosons [6]?

This paper assumes that the Higgs boson is a slepton, but it also explores a new SUSY partnership: What would be required for the 12 gaugino partners of the Standard Model gauge bosons to look like 2 families of 6 left-handed quarks? This pairing up of existing particles would help explain why experiments have not found any new supersymmetric particles.

The elephant in the room for this new partnership is a theorem from Haag, Lopuszanski, and Sohnius (HLS) [7] which showed that supersymmetric partners (in N=1 supersymmetry) have to be in the same representation of the gauge group. This means that if supersymmetry is unbroken, gauge bosons and left-handed quarks cannot be SUSY partners, since they are in different group representations. One way around this is to note that if supersymmetry does indeed exist in Nature, it must be broken anyway, since SUSY partners with the same mass are not readily apparent. In fact, the Minimal Supersymmetric Standard Model [8-10] and similar models always add "soft" SUSY-breaking terms [11,12] by hand. But soft SUSY-breaking terms can be derived from spontaneous SUSY breaking, so it is understood how they arise from an underlying supersymmetric theory. Currently, no spontaneous SUSY-breaking mechanism is known that would lead to a change in gaugino representations.

As a result, the theory presented here cannot be considered supersymmetric. Nonetheless, it leverages the machinery of supersymmetry to facilitate two-loop cancellation of quadratic divergences, thereby lessening the Hierarchy problem. In that sense, the theory belongs to a class of theories that feature cancellation of quadratic divergences without being supersymmetric. For example, many years ago Martinus Veltman proposed a condition on particle masses (especially the top quark and Higgs boson) that leads to cancellation of quadratic divergences at the one-loop level [13]. Later discovery of the top quark and Higgs boson showed that the original Veltman condition is not satisfied at the electroweak scale, although it could potentially be satisfied at the Planck scale [14]. More recently, a non-supersymmetric theory with the same particle content as supersymmetric QED has been constructed such that quadratic divergences are cancelled at the two-loop level [15].

The theory presented here is in a similar vein to that recent work. In this case, a non-supersymmetric theory with the same particle content as supersymmetric SU(3)xSU(3)xU(1)xU(1) is constructed such that quadratic divergences are cancelled at the two-loop level. These cancellations require all four coupling constants to be equal. For that to be the case, the theory should be derivable from a semi-simple Grand Unified Theory (GUT). It is shown that the theory can indeed be derived from SU(3)xSU(3)xSU(3)xSU(3).

Why study this particular theory in the first place? Because after gauge symmetry breaking, the theory reproduces all of the main features of the Standard Model.

To make that connection, a Brout-Englert-Higgs mechanism is proposed that breaks the gauge symmetry down to SU(3)xSU(2)xU(1) such that the remaining two families of six light gauginos are in the (3,2) representation of the group and have the same gauge interactions and charge as two families of Standard Model left-handed quarks. Chiral multiplet fermions included in the theory have the same interactions and charge as the remaining leptons and right-handed quarks of two families of the Standard Model, and the "slepton" scalars play the role of the Higgs boson. Quark masses and Cabbibo mixing arise from gaugino interaction terms. An extension of the theory is also presented that can account for the third family of quarks and leptons.

**Supersymmetric SU(3)xSU(3)xU(1)xU(1):**

Consider a supersymmetric theory $\mathcal{L}_{SUSY}$ involving two SU(3) gauge fields $A_{m\mu} = A_{m\mu}^a t^a$, where $m \in \{1,2\}$ and $t^a$ are SU(3) fundamental representation matrices normalized by $tr(t^a t^b) = \frac{1}{2}\delta^{ab}$. The theory also includes two Abelian gauge fields $B_{3\mu}$ and $B_{8\mu}$. The index notation for the Abelian fields is being used in anticipation of these fields being mapped to the diagonal components of additional SU(3) groups in a Grand Unified Theory (GUT). The supersymmetric partners of the SU(3) gauge fields (the nonAbelian gauginos) are left-handed 2-component fermions denoted by $\lambda_m = \lambda_m^a t^a$. The Abelian gauginos partnered with $B_{3\mu}$ and $B_{8\mu}$ are denoted by $\lambda_3$ and $\lambda_8$. As usual for a supersymmetric theory, the gauginos are in the adjoint representation. As a result, they transform as follows under gauge transformations $\Lambda_m = \Lambda_m^a t^a$ of the two SU(3) groups:

$$\lambda_m \to e^{ig_m \Lambda_m} \lambda_m e^{-ig_m \Lambda_m} \qquad\qquad \lambda_m^T \to e^{-ig_m \Lambda_m^T} \lambda_m^T e^{ig_m \Lambda_m^T}, \tag{1}$$

where the superscript T means to take the transpose of the SU(3) group matrix.

The theory also includes twelve chiral multiplets in the conjugate-fundamental representation, six for each SU(3) group. The scalars and left-handed 2-component fermions of these chiral multiplets are denoted by $\phi_{mR}^{(n)}$ and $\psi_{mR}^{(n)}$, where the index $m$ labels one of the two SU(3) groups, while the indices $(n) \in \{1,2\}$ and $R \in \{1,2,3\}$ specify a "family" and an Abelian charge, respectively.

Covariant derivatives act on the chiral multiplet fields as follows:

$$D_\mu \phi_{mR}^{(n)} = \left(\partial_\mu + ig_m A_{m\mu}^T + \tfrac{1}{\sqrt{2}}(-1)^m ig_0 \left(\tilde{t}_R^3 B_{3\mu} + \tilde{t}_R^8 B_{8\mu}\right)\right)\phi_{mR}^{(n)}, \tag{2}$$

where $g_m$ are the two SU(3) coupling constants and $g_0$ is the coupling for both Abelian fields (the same coupling $g_0$ is used in anticipation of the GUT discussed below). The R-dependent constants $\tilde{t}_R^3$ and $\tilde{t}_R^8$ define the sign and strength of the Abelian interactions and are given by

$$\tilde{t}_1^3 = -\tilde{t}_2^3 = \tfrac{1}{2} \qquad \tilde{t}_3^3 = 0 \qquad \tilde{t}_1^3 = \tilde{t}_2^3 = \tfrac{1}{2\sqrt{3}} \qquad \tilde{t}_3^3 = -\tfrac{1}{\sqrt{3}} \tag{3}$$

One can see that $\tilde{t}_R^3$ and $\tilde{t}_R^8$ have the structure of diagonal SU(3) matrices in the space of the index $R \in \{1,2,3\}$.

The theory $\mathcal{L}_{SUSY}$ is a subgroup of a Grand Unified Theory (GUT) for the semi-simple group SU(3)xSU(3)xSU(3)xSU(3)=SU(3)^4, which is itself a maximal subgroup of the simple group E8. The first two SU(3) groups of SU(3)^4 are the ones described above for $\mathcal{L}_{SUSY}$. Since their group matrices $t_{ij}^a$ have fundamental indices $i \in \{1,2,3\}$, the chiral multiplets described above actually have four sets of indices: $m \in \{1,2\}$, $(n) \in \{1,2\}$, $R \in \{1,2,3\}$, and $i \in \{1,2,3\}$. The $m$=1 multiplets transform in the conjugate representation of the first SU(3) group over indices $i \in \{1,2,3\}$. Within the GUT, they also transform in the conjugate representation of the third SU(3) group over indices $R \in \{1,2,3\}$. The $m$=2 chiral multiplets similarly transform in the conjugate representations of the second and fourth SU(3) groups. Structured this way, each of the four SU(3) groups interacts with the same number of chiral multiplets, so the beta functions for the running couplings of all of the groups are the same.

Symmetry breaking causes the off-diagonal gauge bosons and gauginos of the third and fourth groups to become heavy, along with the diagonal gauge bosons $\tfrac{1}{\sqrt{2}}\left(A_{3\mu}^3 + A_{4\mu}^3\right)$ and $\frac{1}{\sqrt{2}}\left(A^8_{3\mu} + A^8_{4\mu}\right)$ and their gauginos (where lower indices on the gauge fields denote the third and fourth SU(3) groups). The remaining gauge group is SU(3)xSU(3)xU(1)xU(1) with the two Abelian gauge bosons defined by $B_{3\mu} = \frac{1}{\sqrt{2}}\left(-A^3_{3\mu} + A^3_{4\mu}\right)$ and $B_{8\mu} = \frac{1}{\sqrt{2}}\left(-A^8_{3\mu} + A^8_{4\mu}\right)$. The structure of those Abelian gauge bosons reproduces the coupling specified in eq (2). The fact that the Lagrangian $\mathcal{L}_{SUSY}$ can be derived from a supersymmetric GUT is justification for assuming a "unification scale" where all coupling constants are equal.

The supersymmetric Lagrangian $\mathcal{L}_{SUSY}$ for the gauge theory involving the vector fields and chiral multiplets described above is not written down explicitly here, but its form can be deduced from the general SUSY Lagrangians presented in many reviews [16-20] (see for example eq. (5.11) of [16]).

**The SUSY-broken theory:**

Next consider a Lagrangian $\mathcal{L}$ with broken supersymmetry. $\mathcal{L}$ is the same as $\mathcal{L}_{SUSY}$, except for two changes: (i) a change in the gaugino representation and (ii) a change in the sign of the Abelian-scalar coupling (without a change in the Abelian-fermion coupling). To the first point: In $\mathcal{L}$, the 2 Abelian gauginos combine with the 16 nonAbelian ones to transform as 2 families of 9 fermions in the (3,3*) representation of SU(3)xSU(3). In other words, rather than transforming according to eq (1), the "gauginos" in $\mathcal{L}$ transform as follows:

$$\lambda^{(n)} \to e^{ig_1\Lambda_1}\lambda^{(n)}e^{ig_2\Lambda_2^T} \qquad \lambda^{(n)T} \to e^{ig_2\Lambda_2}\lambda^{(n)T}e^{ig_1\Lambda_1^T} \tag{4}$$

where $(n) \in \{1,2\}$ is the same index as the "family" index used for the chiral multiplets.

The changes to the Lagrangian in going from the supersymmetric theory to the nonsupersymmetric one ($\mathcal{L}_{SUSY} \Rightarrow \mathcal{L}$) are limited to the following terms:

$$-i\sum_m \bar{\lambda}^a_m \bar{\sigma}^\mu \partial_\mu \lambda^a_m - i\bar{\lambda}_3 \bar{\sigma}^\mu \partial_\mu \lambda_3 - i\bar{\lambda}_8 \bar{\sigma}^\mu \partial_\mu \lambda_8 \Rightarrow -i\sum_n \bar{\lambda}^{(n)}_{lk} \bar{\sigma}^\mu \partial_\mu \lambda^{(n)}_{kl} \tag{5}$$

$$-2\sum_m \text{tr}\left(\bar{\lambda}_m \bar{\sigma}^\mu \left[g_m A_{m\mu}, \lambda_m\right]\right) \Rightarrow -\sum_n \text{tr}\left(\bar{\lambda}^{(n)} \bar{\sigma}^\mu \left(g_1 A_{1\mu} \lambda^{(n)} + \lambda^{(n)} g_2 A^T_{2\mu}\right)\right) \tag{6}$$

$$-\sqrt{2}i \sum_{n,m,R} \phi^{(n)*}_{mR}\left(g_m \lambda^T_m + (-1)^m \tfrac{1}{\sqrt{2}} g_0 \left(\tilde{t}^3_R \lambda_3 + \tilde{t}^8_R \lambda_8\right)\right)\psi^{(n)}_{mR} + h.c. \Rightarrow$$

$$-i\sum_{m,n,R} \phi^{(n)T}_{mR}\left(g_m\left(\delta_{m1}\lambda^{(n)} + \delta_{m2}\lambda^{(n)T} - \tfrac{1}{3}\lambda^{(n)}_{jj}\right) + \tfrac{1}{3}g_0 \lambda^{(n)}_{jj}\right)\psi^{(n)}_{m'\ne mR} + h.c. , \tag{7}$$

$$D_\mu \phi_{mR}^{(n)} = \left(\partial_\mu + ig_m A_{m\mu}^T + \tfrac{1}{\sqrt{2}}(-1)^m ig_0\left(\tilde{t}_R^3 B_{3\mu} + \tilde{t}_R^8 B_{8\mu}\right)\right)\phi_{mR}^{(n)} \Rightarrow$$

$$D_\mu \phi_{mR}^{(n)} = \left(\partial_\mu + ig_m A_{m\mu}^T - \tfrac{1}{\sqrt{2}}(-1)^m ig_0\left(\tilde{t}_R^3 B_{3\mu} + \tilde{t}_R^8 B_{8\mu}\right)\right)\phi_{mR}^{(n)} \quad (8)$$

where "+h.c." means to add the Hermitian conjugate and the notation conventions of [16] have been used above. The above interactions for $\mathcal{L}$ are gauge invariant under the transformations of eq (4). It should be noted that both $\mathcal{L}_{SUSY}$ and $\mathcal{L}$ are assumed to have "soft" supersymmetry-breaking terms in them. These "soft" terms are exactly the same in both Lagrangians.

The Appendix of this paper shows the following at a unification scale where all coupling constants are equal: a) $\mathcal{L}$ has no quadratic divergences up to two loops, b) the one-loop beta functions for both the gauge coupling and the supersymmetric "d-term" coupling are the same in both $\mathcal{L}_{SUSY}$ and $\mathcal{L}$, and c) differences in the gaugino coupling beta function between the two theories are exactly cancelled in the context of quadratically divergent diagrams.

The Lagrangian $\mathcal{L}$ can also be embedded into an SU(3)^4 GUT, and the findings in the Appendix still hold when comparing the $\mathcal{L}$-version of SU(3)^4 with the $\mathcal{L}_{SUSY}$-version. To see this, one may write the "Abelian" gauginos in $\mathcal{L}$ as $\tfrac{1}{\sqrt{3}}\lambda_{jj}^{(1,2)} = \tfrac{1}{\sqrt{2}}(\lambda_3 \pm i\lambda_8)$. With this notation, the "Abelian" part of eq (7) for $\mathcal{L}$ can be rewritten to exactly mirror $\mathcal{L}_{SUSY}$, except for $\phi_{mR}^{(n)*} \Rightarrow \phi_{mR}^{(n)}$ and global phases $\exp\left(i\varphi_{mR}^{(n=1,2)}\right) = (-1)^m \sqrt{3}\left(\tilde{t}_R^3 \mp i\tilde{t}_R^8\right)$. In other words, the gauginos in the third and fourth SU(3) groups can be constructed to have effectively the same interactions in $\mathcal{L}$ and $\mathcal{L}_{SUSY}$. The one exception is that in $\mathcal{L}$, the "Abelian" gauginos also have an interaction with the $m=1,2$ gauge bosons through eq (6). But that exception is already incorporated into the calculations in the Appendix. The fact that $\mathcal{L}$ can be embedded in a semi-simple GUT justifies the use of the unification scale in the Appendix.

The second part of the paper is devoted to exploring how this model reproduces key features of the Standard Model. It is shown that if certain assumptions are made about gauge symmetry breaking, then the gauginos can be identified with two families of Standard Model left-handed quarks and the sleptons can be identified with Higgs bosons.

**Spontaneous Symmetry Breaking:**

Some additional notation is required to describe the symmetry breaking. First, in order to ensure that the $B_{3\mu}$ acquires a unification-scale mass and to avoid $B_{3\mu}B_8^\mu$ mass cross terms, an R' notation is defined that mixes the R=1 and R=2 fields:

$$\phi^{(n)}_{mR'=1} = \tfrac{1}{\sqrt{2}}\left(\phi^{(n)}_{mR=1} + \phi^{(n)}_{mR=2}\right) \qquad \phi^{(n)}_{mR'=2} = \tfrac{1}{\sqrt{2}}\left(\phi^{(n)}_{mR=1} - \phi^{(n)}_{mR=2}\right) \qquad \phi^{(n)}_{mR'=3} = \phi^{(n)}_{mR=3}. \tag{9}$$

Second, in order to introduce a Cabbibo angle $\theta_C$ for quark mass mixing [21-23], primed fields are defined:

$$\phi'^{(1)}_{mR'=3} = \cos\theta_C \phi^{(1)}_{mR'=3} - \sin\theta_C \phi^{(2)}_{mR'=3}$$

$$\phi'^{(2)}_{mR'=3} = \sin\theta_C \phi^{(1)}_{mR'=3} + \cos\theta_C \phi^{(2)}_{mR'=3}$$

$$\phi'^{(n)}_{mR'=1} = \phi^{(n)}_{mR'=1} \qquad \phi'^{(n)}_{mR'=2} = \phi^{(n)}_{mR'=2}. \tag{10}$$

Although expressed above for scalar fields, these notations are also used for the fermions in chiral multiplets. The *m*=1 SU(3) gauge bosons will be identified as gluons, the *m*=1 chiral fermions (scalars) identified as quarks (squarks), and the *m*=2 chiral fermions (scalars) identified as leptons (sleptons).

Soft SUSY breaking terms are assumed to lead to Brout-Englert-Higgs gauge symmetry breaking that generates the following slepton vacuum expectation values (vevs):

$$\left\langle \phi^{(n)}_{2R'=1} \right\rangle = i\mu_1^{(n)} \begin{pmatrix} 0 \\ (-1)^n \sin\theta_X \\ \cos\theta_X \end{pmatrix} \qquad \left\langle \phi^{(n)}_{2R'=2} \right\rangle = i\mu_2^{(n)} \begin{pmatrix} 0 \\ \cos\theta_X \\ (-1)^n \sin\theta_X \end{pmatrix} \qquad \left\langle \phi'^{(n)}_{2R'=3} \right\rangle = \begin{pmatrix} i\mu_3^{(n)} \\ 0 \\ 0 \end{pmatrix} \tag{11}$$

with

$$\mu_1^{(1)2} - \mu_1^{(2)2} = \mu_2^{(2)2} - \mu_2^{(1)2}. \tag{12}$$

These vevs generate three separate mass terms for gauge bosons: A term proportional to $\cos^2\theta_X \mu_1^{(n)2} + \sin^2\theta_X \mu_2^{(n)2}$, another proportional to $\sin^2\theta_X \mu_1^{(n)2} + \cos^2\theta_X \mu_2^{(n)2}$, and a third proportional to $\mu_3^{(n)2}$. There is no $\sin\theta_X \cos\theta_X$ cross term due to eq (12).

The $\mu_1^{(n)}$ vev is taken to be at a unification scale much higher than the electroweak scale, while $\mu_2^{(n)}$ and $\mu_3^{(n)}$ are below the electroweak scale. Also, it is assumed that that $\sin\theta_X \mu_1^{(n)}$ is at the electroweak scale, so $\sin^2\theta_X \ll 1$ since that scale is much lower than the unification scale.

With these assumptions, the symmetry breaking can be thought to occur in two stages. In the first stage at the unification scale, just the first mass term described above is important, and the symmetry is broken from SU(3)xSU(3)xU(1)xU(1) to the SU(3)xSU(2)xU(1) symmetry of the Standard Model. In the second stage at the electroweak scale, the second two mass terms come into play, and electroweak symmetry breaking occurs.

To describe weak Hypercharge and electric charge for the theory, it is helpful to define the following 6x6 notation for the group matrices:

$$T_1^a = \begin{pmatrix} t^a & 0 \\ 0 & 0 \end{pmatrix} \qquad T_2^a = \begin{pmatrix} 0 & 0 \\ 0 & t^a \end{pmatrix} \qquad T_0 = \frac{1}{\sqrt{12}} \begin{pmatrix} -I & 0 \\ 0 & I \end{pmatrix}, \tag{13}$$

where $I$ represents the 3x3 unit matrix. The upper (lower) 3x3 diagonal block is associated with fields with the index $m=1$ ($m=2$).

The unification-scale symmetry breaking described above imparts large masses to six of the gauge bosons. The gauge bosons that do not become massive at that scale are those with group structures $T_1^a$, $T_2^1$, $T_2^2$, $T_2^3$ and $T^Y = \frac{1}{\sqrt{5}}(2T_0 + T_2^8)$, where the last is a mixture of $B_{8\mu}$ and $A_\mu^8$ fields ($B_{3\mu}$ becomes massive at the unification scale). These 12 gauge bosons correspond to the SU(3) gluons, the SU(2) gauge bosons, and the U(1) Weak hypercharge of the Standard Model. In the notation of eq (13), the group structure of the Weak hypercharge gauge boson is $T^Y = \frac{\sqrt{3}}{2\sqrt{5}} \text{diag}\left(-\frac{2}{3}, -\frac{2}{3}, -\frac{2}{3}, 1, 1, 0\right)$. The coupling constant associated with this field is given by:

$$g_Y^2 = \frac{5g_0^2 g_2^2}{g_0^2 + 8g_2^2} = \frac{5}{9} g^2, \tag{14}$$

where the second equality is true at the unification scale where $g_0 = g_1 = g_2 = g$.

As usual, electroweak symmetry breaking generates masses for the W and Z bosons; only the gluons and the photon remain massless. The group structure of the photon is given by: $T^\gamma = \frac{\sqrt{3}}{\sqrt{8}}\left(\frac{\sqrt{5}}{\sqrt{3}}T^Y + T_2^3\right) = \frac{\sqrt{3}}{\sqrt{8}} \text{diag}\left(-\frac{1}{3}, -\frac{1}{3}, -\frac{1}{3}, 1, 0, 0\right)$. The Weinberg angle for the theory is derived in the usual way from the weak coupling $g_2$ and the weak hypercharge coupling $g_Y$:

$$\sin^2 \theta_W = \frac{\frac{3}{5} g_Y^2}{\frac{3}{5} g_Y^2 + g_2^2}. \tag{15}$$

Since chiral multiplet fields with different values of the R' index have different couplings to the Abelian fields, they have different Weak hypercharges and electric charges. In particular, the group structure for the Weak hypercharge and the photon for different values of the R' index are given by $T_{R'}^Y = \frac{1}{\sqrt{5}}\left(4\sqrt{3}\tilde{t}_{R'}^8 T_0 + T_2^8\right)$ and $T_{R'}^\gamma = \frac{\sqrt{3}}{\sqrt{8}}\left(\frac{\sqrt{5}}{\sqrt{3}} T_{R'}^Y + T_2^3\right)$, respectively. In the conjugate representation, the group structures for the Weak hypercharge boson and the photon are:

$$-T_{R'=1}^Y = -T_{R'=2}^Y = \frac{\sqrt{3}}{2\sqrt{5}} \text{diag}\left(\tfrac{2}{3},\tfrac{2}{3},\tfrac{2}{3},-1,-1,0\right) \qquad -T_{R'=3}^Y = \frac{\sqrt{3}}{2\sqrt{5}} \text{diag}\left(-\tfrac{4}{3},-\tfrac{4}{3},-\tfrac{4}{3},1,1,2\right)$$

$$-T_{R'=1}^\gamma = -T_{R'=2}^\gamma = \frac{\sqrt{3}}{\sqrt{8}} \text{diag}\left(\tfrac{1}{3},\tfrac{1}{3},\tfrac{1}{3},-1,0,0\right) \qquad -T_{R'=3}^\gamma = \frac{\sqrt{3}}{\sqrt{8}} \text{diag}\left(-\tfrac{2}{3},-\tfrac{2}{3},-\tfrac{2}{3},0,1,1\right). \quad (16)$$

The above expressions show that the vevs of eq (11) are only nonzero for electrically neutral components of the scalar fields.

The gauge interactions, Weak hypercharges, and electric charges for the chiral multiplets are consistent with labelling the 2-component left-handed quarks and leptons as follows:

$$\psi_{1R'=1}^{(n)} = \begin{pmatrix} X_1^{c(n)} \\ X_2^{c(n)} \\ X_3^{c(n)} \end{pmatrix} \qquad \psi_{1R'=2}^{(n)} = \begin{pmatrix} d_1^{c(n)} \\ d_2^{c(n)} \\ d_3^{c(n)} \end{pmatrix} \qquad \psi_{1R'=3}^{(n)} = \begin{pmatrix} u_1^{c(n)} \\ u_2^{c(n)} \\ u_3^{c(n)} \end{pmatrix}$$

$$\psi_{2R'=1}^{(n)} = \begin{pmatrix} E^{-(n)} \\ N^{(n)} \\ N''^{(n)} \end{pmatrix} \qquad \psi_{2R'=2}^{(n)} = \begin{pmatrix} e^{-(n)} \\ \nu^{(n)} \\ N'^{(n)} \end{pmatrix} \qquad \psi_{2R'=3}^{(n)} = \begin{pmatrix} \tilde{N}^{(n)} \\ E^{+(n)} \\ e^{+(n)} \end{pmatrix}. \qquad (17)$$

The fields $u_i^{c(n)}$ and $d_i^{c(n)}$ represent two families of antiparticles to right-handed up-type and down-type quarks, respectively. The components $e^{-(n)}$, $e^{+(n)}$ and $\nu^{(n)}$ represent electron-like, positron-like, and neutrino-like leptons, respectively. The remaining leptons and the $X_i^{c(n)}$ quarks will be discussed later.

The gauge interactions and charge of the gauginos are consistent with identifying them as left-handed quarks and labelling their individual components as follows:

$$\lambda^{(n)} = \begin{pmatrix} u_1^{(n)} & d_1'^{(n)} & X_1'^{(n)} \\ u_2^{(n)} & d_2'^{(n)} & X_2'^{(n)} \\ u_3^{(n)} & d_3'^{(n)} & X_3'^{(n)} \end{pmatrix}$$

$$d_i'^{(n)} = \cos\theta_X d_i^{(n)} - (-1)^n \sin\theta_X X_i^{(n)}$$

$$X_i'^{(n)} = \cos\theta_X X_i^{(n)} + (-1)^n \sin\theta_X d_i^{(n)}, \qquad (18)$$

where lower indices represent color. Since $\sin^2\theta_X \ll 1$, the primed components of the gauginos are almost the same as the unprimed components.

The vevs of eq (11) impart masses to the quarks through the following term from eq (7):

$$-ig_2 \sum_n \left\langle \phi_{2R'}^{\prime(n)T} \right\rangle \lambda^{(n)T} \psi_{1R'}^{\prime(n)} + h.c. = g_2 \sum_n \left( \mu_1^{(n)} X_i^{(n)} X_i^{c(n)} + \mu_2^{(n)} d_i^{(n)} d_i^{c(n)} + \mu_3^{(n)} u_i^{(n)} u_i^{\prime c(n)} \right) + h.c., \quad (19)$$

where the $u_i^{\prime c(n)}$ are in the Cabbibo-rotated basis of eq (10). The above mass term has no contribution from the $\lambda_{kk}^{(n)}$ term of eq (7) due to the fact that the symmetry breaking takes place at the unification scale where $g_0 = g_2$. Since the $\mu_1^{(n)}$ are at the unification scale, the masses of the $X_i^{(n)}$ quarks are very large so that they effectively decouple. The $d_i^{(1)}$, $d_i^{(2)}$, $u_i^{(1)}$, $u_i^{(2)}$ are associated with down, strange, up, and charm quarks, respectively. The measured masses for each of these quarks define the values of $\mu_2^{(1)}$, $\mu_2^{(2)}$, $\mu_3^{(1)}$, and $\mu_3^{(2)}$. These quarks experience Cabbibo mass mixing due to the rotated basis introduced in eq (10).

In the electroweak symmetry breaking described above, the W and Z bosons acquire the following masses

$$M_W^2 = \tfrac{1}{2} g_2^2 \sum_n \left( \sin^2\theta_X \mu_1^{(n)2} + \cos^2\theta_X \mu_2^{(n)2} + \mu_3^{(n)2} \right) = \cos^2\theta_W M_Z^2, \quad (20)$$

with the Weinberg angle $\theta_W$ from eq (15). Since the down, strange, up and charm quarks all have masses significantly below that of the W boson, the masses of the W and Z come almost entirely from the $\sin^2\theta_X \mu_1^{(n)2}$ contribution in eq (20). This is why a nonzero $\theta_X$ was introduced in the first place – to allow both the quarks and the gauge bosons to get their masses from slepton vevs.

It is interesting to look at the lepton sector of the theory. After the unification-scale symmetry breaking, it is assumed that the following effective superpotential terms are dynamically generated: $\phi_{2R'=1}^{(n)} I_{33} \phi_{2R'=1}^{(n)}$, $\phi_{2R'=2}^{(n)} I_{33} \phi_{2R'=2}^{(n)}$ and $\phi_{2R'=1}^{(n)} t^2 \phi_{2R'=3}^{(n)}$, where $I_{33} = \text{diag}(0,0,1)$. The first two terms are allowed since the third components of $\phi_{2R'=1,2}^{(n)}$ are SU(2) singlets and have vanishing weak hypercharge (see eq (16)). The last term is allowed due to the fact that SU(2) is a self-conjugate group (through the action of $t^2$) and the first two components of the fields have opposite values for the weak hypercharge. The effective superpotential terms generate the following lepton mass terms: $N''^{(n)}N''^{(n)}$, $N'^{(n)}N'^{(n)}$, $E^{+(n)}E^{-(n)}$ and $N^{(n)}\tilde{N}^{(n)}$. These mass terms (and the analogous ones for their slepton partners) are assumed to be very large but below the unification scale.

After the electroweak-scale symmetry breaking, it is assumed that additional effective superpotential terms proportional to $\phi_{e^+}^{(n)}\phi_{e^-}^{(n)}$ are dynamically generated, where $\phi_{e^\pm}^{(n)}$ are the sleptons partnered with $e^\pm$ leptons. These terms give masses to the electron and muon. Other effective potential terms generated at this stage are assumed to lead to neutrinos mixing with the heavy neutral leptons of the theory.

For calculations below, it is helpful to determine how many of the particles in this theory have unification-scale masses. As mentioned above, soft SUSY-breaking terms lead to unification-scale vevs for the $\phi_{2R'=1}^{(n)}$ sleptons. Six of the twelve degrees of freedom in $\phi_{2R'=1}^{(n)}$ are the "would-be Goldstone bosons" that get "eaten" by the heavy gauge bosons, and the remainder are assumed to have unification-scale masses. After this symmetry breaking, it is assumed that unification-scale mass terms are also generated for all of the squarks as well as for the sleptons partnered with partnered with the $N'$ and $e^{+(n)}$ leptons. These mass terms could be generated by incomplete cancellation of the quadratic divergences below the unification scale and/or by soft SUSY-breaking terms. The remaining "light" scalars (relative to the unification scale) are ones in the SU(2) doublets of $\phi_{2R'=2}^{(n)}$ and $\phi_{2R'=3}^{(n)}$. In this theory, the observed Higgs boson is presumed to be from one of these.

So far it has been shown that after symmetry breaking, the theory can account for two families of quarks and leptons. Before discussing the third family, it is interesting to look at implications of the Weinberg angle and renormalization group scaling.

From eqs (14) and (15), one can see that $\sin^2\theta_W = \frac{1}{4}$ at the unification scale. The measured value of ~0.23 at the electroweak scale is very close to this. That would imply that the relation $g_Y^2 = \frac{5}{9}g_2^2$ valid at the unification scale should also approximately hold at the electroweak scale. For that to be the case, $g_Y^2$ and $g_2^2$ would have to have similar renormalization group scale dependence from the unification scale down to the electroweak scale. The one-loop $\beta$ functions for these couplings take the usual form (e.g. [20]):

$$\beta_G = \frac{g_G^3}{16\pi^2} b_G$$

$$b_2 = -\tfrac{11}{3}(2) + \tfrac{1}{3} n_f + \tfrac{1}{6} n_s$$

$$b_Y = \tfrac{3}{20}\left( \tfrac{2}{3}\sum_f y_f^2 + \tfrac{1}{3}\sum_s y_s^2 \right), \tag{21}$$

where $n_f$ and $n_s$ are the numbers of "light" (relative to the unification scale) fermion and scalar SU(2) doublets, and $y_f$ and $y_s$ are the weak hypercharges from eq (16) for each "light" fermion and scalar. For $g_Y^2$ and $g_2^2$ to have similar one-loop renormalization group scale dependence, the following condition would have to be met:

$$b_2 \sim \tfrac{5}{9} b_Y. \tag{22}$$

As discussed above, the following fields are assumed to have unification-scale masses: $X^{(n)}$, $X^{c(n)}$, $\phi_{1R'}^{(n)}$, $\phi_{2R'=1}^{(n)}$, $\phi_{N'}^{(n)}$, $\phi_{e^+}^{(n)}$. Assuming all other fields are "light", one finds $b_2 = -\tfrac{8}{3}$ and $\tfrac{5}{9} b_Y = \tfrac{58}{27}$. These beta functions are not very similar. But adding more chiral multiplets will bring them closer and can also accommodate a third family.

**Addition of a third family:**

A new index $(p) \in \{1, 2\}$ will be added to the chiral multiplet fields, changing them to $\phi_{mR}^{(n)(p)}$ and $\psi_{mR}^{(n)(p)}$, effectively doubling their number. Gauginos with index $(n)$ still interact with chiral multiplets having the same index $(n)$. From the gaugino standpoint, the new index $(p)$ just goes along for the ride. Since $(p)$ also goes along for the ride in $\mathcal{L}_{SUSY}$ gaugino interactions, this doubling of chiral multiplets does not change the conclusions of the Appendix.

In addition, four more chiral multiplets are added: two $\left(\Phi^{(n)}, \Lambda^{(n)}\right)$ that transform in the (3,3*) representation and two $\left(\tilde{\phi}_m, \tilde{\psi}_m\right)$ that transform in the adjoint+singlet representations of the two SU(3) groups. Specifically, the fermion fields transform as follows:

$$\Lambda^{(n)} \to e^{ig_1 \Lambda_1} \Lambda^{(n)} e^{ig_2 \Lambda_2^T} \qquad \Lambda^{(n)T} \to e^{ig_2 \Lambda_2} \Lambda^{(n)T} e^{ig_1 \Lambda_1^T}$$

$$\tilde{\psi}_m \to e^{ig_m \Lambda_m} \tilde{\psi}_m e^{-ig_m \Lambda_m} \qquad \tilde{\psi}_m^T \to e^{-ig_m \Lambda_m^T} \tilde{\psi}_m^T e^{ig_m \Lambda_m^T}, \tag{22}$$

with the same transformations for their scalar partners. Following arguments similar to those in the Appendix, gauge invariant interactions of these fields with gauginos can be defined for $\mathcal{L}$ that mirror analogous ones in $\mathcal{L}_{SUSY}$, such that the ones in $\mathcal{L}$ do not generate one-loop quadratic divergences at the unification scale. Addition of these new fields can also be accommodated in the SU(3)^4 GUT discussed above by adding similar fields to the other two SU(3) groups. As an added bonus, the additional $\tilde{\phi}_{3,4}^{(n)}$ scalars would be in the correct representations to facilitate the symmetry breaking from SU(3)^4 to SU(3)xSU(3)xU(1)xU(1) that was described previously.

The superpotential for the extended theories (both $\mathcal{L}_{SUSY}$ and $\mathcal{L}$) include the following gauge invariant and holomorphic terms: $tr\left(\tilde{\phi}_m^{(n)}\tilde{\phi}_m^{(n)}\right)$ and $-ic_{R'}^{(n)}\phi_{1R'}^{(n)(2)T}\Phi^{(n)}\phi_{2R'}^{(n)(2)}$, where $c_{R'}^{(n)}$ are superpotential coupling constants. The first term generates unification-scale masses for the $\left(\tilde{\phi}_m^{(n)},\tilde{\psi}_m^{(n)}\right)$ fields. The second term generates couplings analogous to those of eq (7) with $\Lambda^{(n)}$ playing the role of the third and fourth left-handed quark families, while $\psi_{1R'}^{(n)(2)}$ and $\psi_{2R'}^{(n)(2)}$ play the roles of third and fourth families of right-handed quarks and leptons. In this model, $\Lambda^{(n)}$ does not interact with the two (p)=1 right-handed quark and lepton families, so the (n)=1,2 (p)=1 families can be identified with first two families discussed previously before addition of the new chiral multiplets.

For the third and fourth families, it is also assumed that the sleptons acquire vacuum expectation values analogous to those in eq (11). Following arguments similar to those above, unification-scale masses are generated for $X^{(n)(p)}$, $X^{c(n)(p)}$, $\phi_{1R'}^{(n)(p)}$, $\phi_{N'}^{(n)(p)}$ and $\phi_{e^+}^{(n)(p)}$. Also, large (but below unification scale) mass terms are generated for $E^{+(n)(p)}E^{-(n)(p)}$ and $N^{(n)(p)}\tilde{N}^{(n)(p)}$ and their slepton partners. The slepton vevs and superpotential couplings can be adjusted such that the remaining quarks and leptons for (p)=2 have one family with masses matching those of the bottom/top/tau and another family with masses heavier than current experimental measurements.

A key difference for the third and fourth family of right-handed quarks is that they interact not only with the heavier left-handed quarks $\Lambda^{(n)}$, they also interact with the gauginos $\lambda^{(n)}$ through a generalized eq (7). The fact that both interactions occur generates mass mixing between the quark families. The slepton vevs can be chosen such that they do a good job of reproducing the measured Cabibbo-Kobayashi–Maskawa (CKM) mass matrix [21-23]. For example, the slepton field combination to get a vev associated with the bottom quark can be proportional to $\phi_{2R'=2}^{(2)(2)} + (\rho - i\eta)\phi_{2R'=2}^{(1)(2)}$, where $\rho$ and $\eta$ are Wolfenstein parameters. After tuning the vevs to get the CKM matrix correct, the superpotential couplings $c_{R'}^{(n)}$ can be chosen to get the masses correct.

The modified theory with additional chiral multiplets presented above results in the following values for the beta functions after unification-scale symmetry breaking:

$$\tfrac{5}{9}b_Y = \tfrac{13}{3} \quad \text{for U(1)}$$
$$b_2 = 3 \quad \text{for SU(2)}$$
$$b_3 = -5 \quad \text{for SU(3)}. \tag{23}$$

By comparing $b_2$ and $b_3$, one can derive a unification scale of $\sim 10^9$ GeV, many orders of magnitude lower than in most GUTs. Using that scale and the fact that $g_Y^2 = \tfrac{5}{9}g_2^2$ at the unification scale results in $\sin^2\theta_W(M_Z) \sim 0.22$, in reasonable agreement with measured results for a one-loop calculation.

The reason that the unification scale for this model is so much lower than that of SU(5) or many other GUTs is because this model features $g_Y^2 = \tfrac{5}{9}g_2^2$ at the unification scale rather than them being equal. Said another way, $\sin^2\theta_W = \tfrac{1}{4}$ at the unification scale, rather than $\tfrac{3}{8}$. The net result is that one does not have to go to as high of an energy scale to achieve unification.

**Conclusion:**

This paper has shown that it is possible to take the particle content of a supersymmetric SU(3)xSU(3)xU(1)xU(1) gauge theory and construct a non-supersymmetric theory that has the following features: (i) The gauginos of the theory can be identified as two families of left-handed quarks, (ii) sleptons of the theory can be identified as Higgs bosons, (iii) the theory can be derived from an SU(3)xSU(3)xSU(3)xSU(3) Grand Unified Theory, (iv) at a unification scale of $\sim 10^9$ GeV, two-loop quadratic divergences are cancelled in the theory, and (v) after unification-scale symmetry breaking, the theory reproduces the main features of the Standard Model.

**Appendix:**

This Appendix shows the following about the SUSY-broken theory $\mathcal{L}$ compared to the supersymmetric theory $\mathcal{L}_{SUSY}$ in the Landau gauge and at a unification scale where all coupling constants are equal: (a) both theories have the same one-loop wave function renormalization constants for all fields, (b) both theories have the same one-loop beta functions for their gauge couplings and for their supersymmetric d-term couplings, (c) a difference between the theories in the one-loop gaugino vertex is cancelled in quadratically divergent diagrams, and (d) two-loop quadratically divergent diagrams are the same in both theories. Since $\mathcal{L}_{SUSY}$ is free of quadratic divergences, it means that $\mathcal{L}$ is also free of them to two loops at the unification scale. The analysis is simplified by the fact that $\mathcal{L}$ is constructed to be exactly the same as $\mathcal{L}_{SUSY}$ except for terms involving gauginos. As a result, it is sufficient to just compare diagrams involving gauginos.

From a Feynman graph standpoint, the difference between the supersymmetric and SUSY-broken theories only shows up in the gaugino propagator and vertices:

$$\begin{array}{c}
\text{(propagator)} \quad \delta^{ab}; \delta_{33}; \delta_{88} \quad \xrightarrow{\mathcal{L}_{SUSY} \Rightarrow \mathcal{L}} \quad \delta_{ik}\delta_{jl} \\[1em]
\text{(gaugino-gaugino-gauge)} \quad g_m f^{cba} \quad \xrightarrow{\mathcal{L}_{SUSY} \Rightarrow \mathcal{L}} \quad -ig_m\left(\delta^{m1} t^c_{ik}\delta_{jl} + \delta^{m2} t^c_{jl}\delta_{ik}\right) \\[1em]
\left(\sqrt{2}g_m t^a_{ij}\right); g_0\left(\tilde{t}^3_R; \tilde{t}^8_R\right)\delta_{ij}\left(-\delta_{m1}+\delta_{m2}\right) \quad \xrightarrow{\mathcal{L}_{SUSY} \Rightarrow \mathcal{L}} \quad \left(\delta_{m1}\hat{g}_1\delta_{jk}\delta_{il} + \delta_{m2}\hat{g}_2\delta_{jl}\delta_{ik} - \tfrac{1}{3}\left(\hat{g}_m - \hat{g}_0\right)\delta_{ij}\delta_{kl}\right)\delta_{m'\neq m}
\end{array}$$

(A1)

In these diagrams (and the ones below), scalars, fermions (from chiral multiplets), gauge bosons, and gauginos are represented by dashed lines, solid lines, wavy lines, and solid+wavy lines, respectively. The arrows point from $\phi, \lambda, \psi$ and toward $\phi^*, \bar{\lambda}, \bar{\psi}$ and just the group structure is shown. For each of the diagrams showing a gaugino coupled to a scalar and fermion, there is another diagram (not shown) that has the arrows reversed. Hats have been added to the gaugino coupling constants of $\mathcal{L}$ to differentiate them from the gauge coupling constants. The reason to differentiate is because in $\mathcal{L}$ with broken supersymmetry, there is no guarantee that the running gaugino coupling will remain equal to the running gauge coupling as the scale is changed. That being said, it will be shown that before unification-scale symmetry breaking, they are indeed the same.

**One-loop quadratic divergence cancellation and wave function renormalization:**

In both theories, there are four potentially quadratically divergent one-loop diagrams:

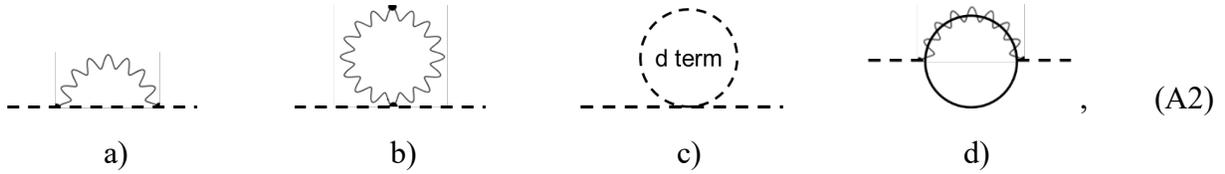

$$\text{a)} \qquad \text{b)} \qquad \text{c)} \qquad \text{d)} \qquad \text{(A2)}$$

where diagram c) involves the supersymmetric 4-scalar interaction term (the d-term) that comes from solving the equations of motion for the auxiliary d field in the Lagrangian. In this Appendix, the Landau gauge is assumed. As a result, diagram a) does not produce a quadratic divergence. It does produce a logarithmic divergence that contributes to the scalar wave function renormalization constant, but that contribution is the same in both theories by construction since the diagram has no gauginos. Similarly, diagrams b) and c) also produce the same result in both theories since they do not have gauginos. Since in $\mathcal{L}_{SUSY}$, the diagrams of (A2) all cancel, they will also cancel in $\mathcal{L}$ if diagram d) produces the same result in both theories.

Diagram d) is shown again below, along with the results produced from each theory.

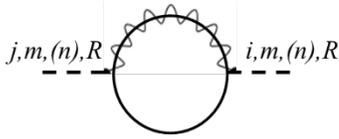

For $\mathcal{L}_{SUSY}$: $\qquad \left( 2g_m^2 \left(t^a t^a\right)_{ij} + g_0^2 \left( \left(\tilde{t}_R^3\right)^2 + \left(\tilde{t}_R^8\right)^2 \right) \delta_{ij} \right) \cdots$

$$= \left( \tfrac{8}{3} g_m^2 + \tfrac{1}{3} g_0^2 \right) \delta_{ij} \cdots$$

For $\mathcal{L}$ m=1: $\qquad \left( \hat{g}_1^2 \left( \delta_{ik}\delta_{pl} - \tfrac{1}{3}\delta_{ip}\delta_{kl} \right)\left( \delta_{jk}\delta_{pl} - \tfrac{1}{3}\delta_{jp}\delta_{kl} \right) + \tfrac{1}{9}\hat{g}_0^2 \delta_{ip}\delta_{kl}\delta_{jp}\delta_{kl} \right) \cdots$

$$= \left( \tfrac{8}{3} \hat{g}_1^2 + \tfrac{1}{3} \hat{g}_0^2 \right) \delta_{ij} \cdots$$

For $\mathcal{L}$ m=2: $\quad \left(\frac{8}{3}\hat{g}_2^2 + \frac{1}{3}\hat{g}_0^2\right)\delta_{ij} \cdots$ (A3)

In other words, at the unification scale where all coupling constants are the same (including unhatted and hatted), both theories give the same contribution to the one-loop diagram above (where "$\cdots$" stands for momentum and spin dependence which is the same in both theories). As described previously, it follows that $\mathcal{L}$ has no one-loop quadratic divergences at the unification scale. Furthermore, the one-loop scalar wave function renormalization constants for $\mathcal{L}$ and $\mathcal{L}_{SUSY}$ are also the same.

By a similar calculation as above, results for the following diagram can be calculated.

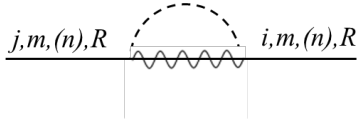

For $\mathcal{L}_{SUSY}$: $\quad \left(\frac{8}{3}g_m^2 + \frac{1}{3}g_0^2\right)\delta_{ij} \cdots$

For $\mathcal{L}$: $\quad \left(\frac{8}{3}\hat{g}_{m'}^2 + \frac{1}{3}\hat{g}_0^2\right)_{m'\neq m}\delta_{ij} \cdots$ (A4)

Again, at the unification scale, both theories produce the same correction. Since all other corrections are also the same by construction, the fermion wave function renormalization constants for $\mathcal{L}$ and $\mathcal{L}_{SUSY}$ are the same.

The one-gaugino-loop correction to the gauge boson propagator is depicted by:

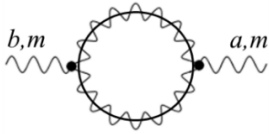

For $\mathcal{L}_{SUSY}$: $\quad -g_m^2 f^{acd} f^{bdc} \cdots = 3g_m^2 \delta^{ab} \cdots$

For $\mathcal{L}$: $\quad \delta_{jl}\delta_{jl}g_m^2 \sum_n tr\left(t^a t^b\right) \cdots = 3g_m^2 \delta^{ab} \cdots$. (A5)

Both theories again produce the same correction. Furthermore, there is no one-loop correction to the Abelian gauge bosons that involves gauginos, so the wave function renormalization constants for $\mathcal{L}$ and $\mathcal{L}_{SUSY}$ are the same for all of the gauge bosons.

A one-loop correction to the gaugino propagator is depicted by:

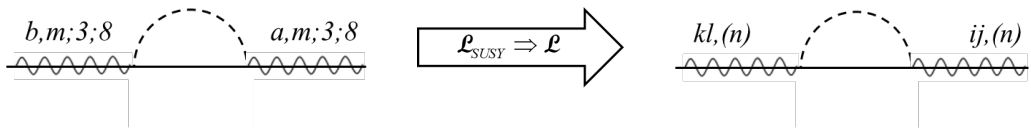

For $\mathcal{L}_{SUSY}$ nonAbelian gauginos: $\sum_{n,R} 2g_m^2 tr(t^a t^b)\cdots = \delta^{ab} g_m^2 \sum_{n,R}$

For $\mathcal{L}_{SUSY}$ Abelian gauginos: $g_0^2 \sum_{n,R}$

For $\mathcal{L}$ traceless gauginos: $\sum_{m,R} \hat{g}_m^2 \left(\delta_{ip}\delta_{jq} - \tfrac{1}{3}\delta_{ij}\delta_{pq}\right)\left(g_m \delta_{pk}\delta_{ql} - \tfrac{1}{3}\delta_{pq}\delta_{kl}\right)\cdots$

$$= \left(\delta_{ik}\delta_{jl} - \tfrac{1}{3}\delta_{ij}\delta_{kl}\right) \sum_{m,R} \hat{g}_m^2 \cdots$$

For $\mathcal{L}$ "Abelian" gauginos: $\tfrac{1}{3}\delta_{ij}\delta_{kl} \hat{g}_0^2 \sum_{m,R}$  (A6)

At the unification scale, all of the above corrections to gaugino propagators are the same.

There is another one-loop correction to the gaugino propagator involving a gauge boson loop. In the Landau gauge that correction is finite. This section is only considering divergent wave function renormalization constants (assuming a minimal renormalization scheme), so that correction is not addressed here. However, it is addressed in (A12) below. Eq (A6) has shown that the gaugino wave function renormalization constant is the same in both theories at the unification scale.

More generally, it has now been shown that the one-loop wave function renormalization constants for all fields are the same in both $\mathcal{L}_{SUSY}$ and $\mathcal{L}$ at the unification scale.

**One-loop vertex corrections:**

In $\mathcal{L}_{SUSY}$, the cancellation of diagrams in (A2) persists as the scale is changed. This is due to the fact that supersymmetry ensures that the couplings involved in diagrams c and d (the "d-term coupling" and the "gaugino coupling") have the same scale dependence as the gauge couplings in diagrams a and b. For $\mathcal{L}$, supersymmetry is broken, so it is not assured that the d-term and gaugino couplings will have the same scale dependence as the gauge coupling. This section will show two things at the unification scale. First, the d-term coupling does indeed have the same one-loop scale dependence as the gauge coupling. Second, a difference between $\mathcal{L}_{SUSY}$ and $\mathcal{L}$ in the scale dependence of the gaugino coupling cancels in the context of quadratically divergent diagrams.

Since both $\mathcal{L}_{SUSY}$ and $\mathcal{L}$ are gauge theories, gauge invariance ensures that for nonAbelian fields (m=1,2), the gauge couplings to scalars in diagrams a) and b) of (A2) have the same beta function as the 3-gauge coupling. The one-gaugino-loop correction to the latter is depicted by the following diagram:

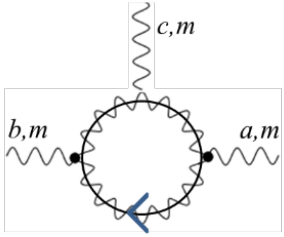

For $\mathcal{L}_{SUSY}$: $g_m^3 f^{ade} f^{bef} f^{cfd} \cdots = \left(\frac{3}{2} g_m^2\right)\left(g_m f^{abc}\right)\cdots$

For $\mathcal{L}$: $ig_m^3 \sum_{n=1,2} \delta_{jl} \delta_{jl} tr\left(t^a t^c t^b\right) \cdots = \left(\frac{3}{2} g_m^2\right)\left(g_m f^{abc}\right)\cdots$ (A7)

Both Lagrangians have the same one-loop contribution. Since all other one-loop corrections to the 3-gauge vertex are the same in both theories by construction, and since eq (A5) shows that gauge wave function renormalization constants are the same in both theories, it follows that the gauge coupling constants $g_1$ and $g_2$ have the same one-loop beta functions in both $\mathcal{L}_{SUSY}$ and $\mathcal{L}$.

The one-loop beta function for the Abelian gauge coupling $g_0$ requires evaluation of the following diagram:

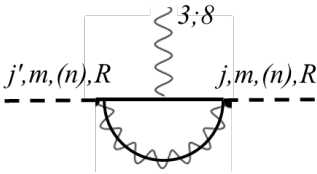

For $\mathcal{L}_{SUSY}$: $g_0\left(i\frac{1}{\sqrt{2}}(-1)^m \tilde{t}_R^{3;8}\right)\left(\frac{8}{3}g_m^2 + \frac{1}{3}g_0^2\right)\delta_{jj'}\cdots$

For $\mathcal{L}$: $g_0\left(-i\frac{1}{\sqrt{2}}(-1)^m \tilde{t}_R^{3;8}\right)\left(\frac{8}{3}\hat{g}_m^2 + \frac{1}{3}\hat{g}_0^2\right)\delta_{jj'}\cdots$ (A8)

Diagram (A8) shows coupling of Abelian gauge fields to fermions which has the same factor of $(-1)^m$ in $\mathcal{L}_{SUSY}$ and $\mathcal{L}$. The minus sign for $\mathcal{L}$ is due to the fact that in $\mathcal{L}$, the fermion in the loop has $m' \neq m$. But the theories have opposite signs for their couplings to scalars (see eq (8)), so the vertex renormalization constant is the same in both theories. Since the wave function renormalization constants for all fields involved in the vertex are also the same in both theories, and all other relevant diagrams are the same by construction, the Abelian gauge coupling beta functions and scaling behavior are the same in both theories at the unification scale.

The one-gaugino-loop correction to the four-scalar d-term vertex is represented by the following diagram:

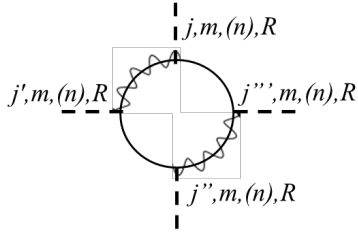

For $\mathcal{L}_{SUSY}$: $\delta_{i'i''}\delta_{i''i}\left(2t_{ij}^c t_{j'i'}^c g_m^2 + \left(\left(\tilde{t}_R^3\right)^2 + \left(\tilde{t}_R^8\right)^2\right)g_0^2 \delta_{ij}\delta_{i'j'}\right)\left(2t_{i''j'}^b t_{j''i''}^b g_m^2 + \left(\left(\tilde{t}_R^3\right)^2 + \left(\tilde{t}_R^8\right)^2\right)g_0^2 \delta_{i''j'}\delta_{j''i''}\right)\cdots$

$= \left(\left(\delta_{ii'}\delta_{jj'} - \tfrac{1}{3}\delta_{ij}\delta_{i'j'}\right)g_m^2 + \tfrac{1}{3}g_0^2 \delta_{ij}\delta_{i'j'}\right)\left(\left(\delta_{i''i}\delta_{j''j''} - \tfrac{1}{3}\delta_{i'j'}\delta_{j''i}\right)g_m^2 + \tfrac{1}{3}g_0^2 \delta_{i'j'}\delta_{j''i}\right)\cdots$

For $\mathcal{L}$: $\left(\left(\delta_{ii'}\delta_{jj'} - \tfrac{1}{3}\delta_{ij}\delta_{i'j'}\right)\hat{g}_m^2 + \tfrac{1}{3}\hat{g}_0^2 \delta_{ij}\delta_{i'j'}\right)\left(\left(\delta_{i''i}\delta_{j''j''} - \tfrac{1}{3}\delta_{i'j'}\delta_{j''i}\right)\hat{g}_m^2 + \tfrac{1}{3}\hat{g}_0^2 \delta_{i'j'}\delta_{j''i}\right).$ (A9)

At the unification scale, both theories produce the same one-loop correction. Since all other one-loop corrections to the d-term vertex are the same in both theories by construction, and since eq (A3) shows that scalar wave function renormalization constants are the same, it follows that the d-term coupling constants have the same one-loop beta functions and one-loop scaling behavior in both theories at the unification scale. As a consequence, at the unification scale, the one-loop beta functions and scaling behavior of the d-term couplings in $\mathcal{L}$ are the same as those of the gauge couplings, since both are equal to their $\mathcal{L}_{SUSY}$ counterparts.

In the Landau gauge, the only divergent one-loop correction to the gaugino coupling vertex is given by the diagram below:

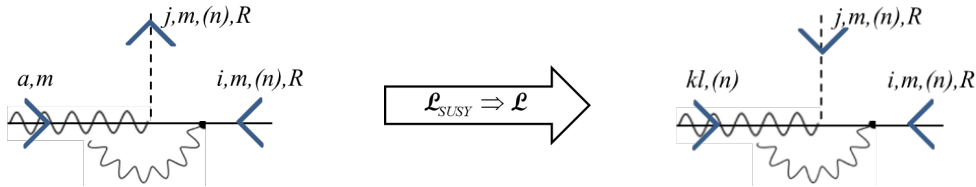

For $\mathcal{L}_{SUSY}$: $\sqrt{2}g_m t_{ij}^b \left(g_m f^{cab}\right)\left(ig_m t_{i'i}^{Tc}\right)\cdots = \sqrt{2}g_m t_{ij}^a \left(\tfrac{3}{2}g_m^2\right)\cdots$

For $\mathcal{L}$ m=1: $\left(\hat{g}_1 \left(\delta_{jk'}\delta_{i'l'} - \tfrac{1}{3}\delta_{ij}\delta_{k'l'}\right) + \tfrac{1}{3}\hat{g}_0 \delta_{ij}\delta_{k'l'}\right)\left(-ig_2 t_{l'l}^c \delta_{k'k}\right)\left(ig_2 t_{i'i}^{Tc}\right)\cdots$

$$= \left( \left( \delta_{jk}\delta_{il} - \tfrac{1}{3}\delta_{ij}\delta_{kl} \right)\left( \tfrac{7}{6}\hat{g}_1 + \tfrac{1}{6}\hat{g}_0 \right) + \tfrac{1}{3}\delta_{ij}\delta_{kl} \tfrac{4}{3}\hat{g}_1 \right) g_2^2 \cdots$$

For $\mathcal{L}$ m=2: $\left( \left( \delta_{jk}\delta_{il} - \tfrac{1}{3}\delta_{ij}\delta_{kl} \right)\left( \tfrac{7}{6}\hat{g}_2 + \tfrac{1}{6}\hat{g}_0 \right) + \tfrac{1}{3}\delta_{ij}\delta_{kl} \tfrac{4}{3}\hat{g}_2 \right) g_1^2 \cdots$. (A10)

There is no one-loop correction with a scalar in the loop since it is not possible to draw such a diagram in which the arrows of the last vertex in (A1) are all consistent. At the unification scale, the one-loop gaugino vertex correction for $\mathcal{L}$ is $\tfrac{8}{9}$ that of $\mathcal{L}_{SUSY}$. But the vertex correction for $\mathcal{L}_{SUSY}$ only applies to vertices involving the 8 nonAbelian gauginos in each family, whereas the correction for $\mathcal{L}$ applies to vertices involving all 9 of the gauginos. As a result, in any diagram where gauginos appear only in loops, the vertex correction for $\mathcal{L}_{SUSY}$ will be applied $\tfrac{8}{9}$ of the times that it will be applied for $\mathcal{L}$. In other words, $\mathcal{L}_{SUSY}$ and $\mathcal{L}$ give the same result for these one-loop vertex corrections in the context of any diagram that has no external gaugino lines. Since no quadratically divergent diagrams have external gaugino lines, in the context of those diagrams, the one-loop beta function and scaling behavior of the gaugino coupling is the same in both theories.

More generally, it has now been shown that in the context of diagrams with no external gaugino lines, the one-loop beta functions for all coupling constants are the same in both $\mathcal{L}_{SUSY}$ and $\mathcal{L}$ at the unification scale.

**Two-loop quadratic divergence cancellation:**

There are three types of quadratically divergent two-loop diagrams: (i) diagrams with no gauginos, (ii) diagrams that modify those of (A2) by incorporating the one-loop wave function or vertex corrections considered previously in this Appendix, and (iii) other diagrams. Type (i) diagrams give the same result in $\mathcal{L}_{SUSY}$ and $\mathcal{L}$ by construction. The previous sections of this Appendix have shown that type (ii) diagrams also give the same result.

The only diagram in the "other" category is the following one:

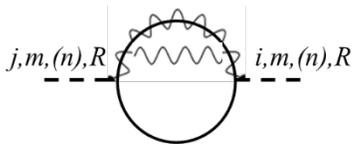

(A11)

This diagram involves a one-loop correction to the gaugino propagator that was not considered previously since it is not divergent in the Landau gauge. However, the finite propagator correction still leads to a quadratic divergence in the above diagram.

The relevant gaugino propagator correction is depicted by:

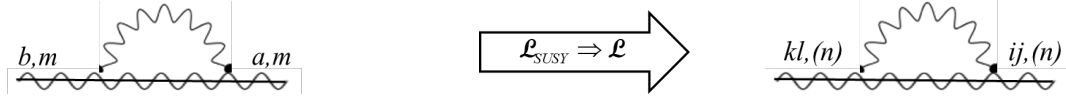

For $\mathcal{L}_{SUSY}$: $\quad -g_m^2 f^{acd} f^{bdc} \cdots = 3g_m^2 \delta^{ab} \cdots$

For $\mathcal{L}$: $\quad \left(g_1^2 \left(t^a t^b\right)_{ik} \delta_{jl} + g_2^2 \left(t^{aT} t^{bT}\right)_{jl} \delta_{ik}\right) \cdots = \frac{4}{3} \delta_{ik} \delta_{jl} \sum_m g_m^2 \cdots.$ (A12)

Just as in (A10), the correction for $\mathcal{L}$ at the unification scale is $\frac{8}{9}$ the one for $\mathcal{L}_{SUSY}$. But the correction for $\mathcal{L}_{SUSY}$ only applies to nonAbelian gauginos, whereas the one for $\mathcal{L}$ applies to all gauginos. This means that $\mathcal{L}_{SUSY}$ and $\mathcal{L}$ generate the same correction for diagram (A11). More generally, it has been shown that $\mathcal{L}_{SUSY}$ and $\mathcal{L}$ generate the same result for all quadratically divergent two-loop diagrams. Since all quadratic divergences cancel in $\mathcal{L}_{SUSY}$, all two-loop quadratic divergences cancel in $\mathcal{L}$ at the unification scale.